\documentclass[nofootinbib,showpacs,showkeys]{revtex4-2}
\usepackage{amssymb}
\usepackage{amsmath}
\usepackage{bm}
\usepackage{hyperref}
\usepackage{latexsym}
\usepackage{dcolumn}
\usepackage{float}
\usepackage{hyperref}
\begin{document}

\title{Brans-Dicke unimodular gravity}
\author{Alexandre M.R. Almeida}%
 \email{alexandreRalmeida@hotmail.com}
\affiliation{N\'ucleo Cosmo-ufes \& Departamento de F\'isica, UFES, Vit\'oria, ES, Brazil}
\author{Júlio C. Fabris}
\email{julio.fabris@cosmo-ufes.org}
\affiliation{N\'ucleo Cosmo-ufes \& Departamento de F\'isica, UFES, Vit\'oria, ES, Brazil}
\affiliation{National Research Nuclear University MEPhI, Kashirskoe sh. 31, Moscow 115409, Russia}
\author{Mahamadou Hamani Daouda}
\email{daoudah77@gmail.com}
\affiliation{Departement de Physique \&  Faculté des Sciences et Techniques (FAST) \& Université Abdou Moumouni de Niamey (UAM)
BP: 10662 Niamey, Niger}
\author{Richard Kerner}
\email{richard.kerner@sorbonne-universite.fr}
\affiliation{LPTMC, Sorbonne-Université - CNRS UMR 7600, Paris, France}
\author{Hermano Velten}%
 \email{hermano.velten@ufop.edu.br}
\affiliation{Departamento de F\'isica, Universidade Federal de Ouro Preto (UFOP), Campus Morro do Cruzeiro, 35400-000, Ouro Preto-MG Brazil}
\author{W. S. Hip\'olito-Ricaldi}
\email{wiliam.ricaldi@ufes.br}
\affiliation{N\'ucleo Cosmo-ufes \& Departamento de Ciências Naturais, UFES, São Mateus, ES, Brazil}
\date{\today}

\begin{abstract}We propose a unimodular version of the Brans-Dicke theory designed with a  constrained Lagrangian formulation. The resulting field equations are traceless. The vacuum solutions in the cosmological background reproduce the corresponding solutions of the usual Brans-Dicke theory but with a cosmological constant term. A perturbative analysis of the scalar modes is performed and stable and unstable configurations appear in contrast with the Brans-Dicke case for which only stable configurations occur. On the other hand, tensorial modes in this theory remains the same as in the traditional Brans-Dicke theory. 
\end{abstract}

\maketitle

\section{Introduction} \label{sec:introduction}

The unimodular gravity is a possible alternative formulation of a geometric theory of gravity that has appeared soon after the General Relativity (GR) theory \cite{pedra}. In principle,  it has the same content as the GR theory except for one important difference: an
extra constraint fixing the determinant of the metric  $g_{\mu\nu}$ equal to one. This condition fixes a class of allowed coordinate systems to be used. More precisely, this condition restricts the general diffeomorphism (Diff) invariance,  on which the GR theory reposes, to a subclass of transformation, called transverse 
diffeomorphism (TDiff) \cite{trans}. The general Diff contains 10 parameters, while the unimodularity condition implies that the TDiff has 9 parameters. The unimodular field equations involve traceless tensors on both sides, a feature that reflects the absence of some informations contained in the trace of the field equations. Even though, under an important extra hypothesis, the conservation of the energy-momentum tensor, the GR equations can be recovered but with an integration constant which has the form of a cosmological constant term. This property of the unimodular theory has been considered as an attractive one since it could alleviate the cosmological constant problem. Some authors have even argued that the content of the unimodular theory is indeed the same as GR, and that the general Diff can be recovered, see for example Ref. \cite{wei}. Even if this may occur classically, at quantum level the unimodular theory may be quite different from GR. For a recent review on this subject, see Ref. \cite{ant,esp}.

The vacuum solutions in unimodular theory are in principle the same as the corresponding GR ones. If the unimodular condition is imposed, the Schwarzschild solution is written in a very peculiar system of coordinates, but this restriction can be overcome by easing the unimodular condition via introduction of an arbitrary external field. In the presence of matter, however, such connection depends on whether the conservation of the energy-momentum tensor is imposed or not. If the energy-momentum tensor is conserved, the resulting class of solutions is the same as in GR with a cosmological constant
added.  If the energy-momentum tensor does not obey the usual conservation laws, the correspondence with GR may be
broken. In cosmology, for example, the resulting system of equations is underdetermined, the evolution of the universe being sensitive only to the combination $\rho + p$ (which is related to the enthalpy), and some additional hypothesis must be imposed in order to
close the system and to obtain specific solutions. These aspects have been discussed in Refs. \cite{alvarez,velten}. It must be remarked that even imposing the conservation of the energy-momentum tensor, some new features appear at perturbative level, see Ref. \cite{brand}.

To our present knowledge, no generalization of the unimodular theory to a scalar-tensor framework is available, like for example a unimodular version of the Brans-Dicke theory. To explore such generalizations is the main goal of the present work. Some analysis using scalar field has been made previously but without a full unimodular Brans-Dicke formulation \cite{bohmer,esp}. Using a constrained action, by introducing a Lagrange multiplier (a procedure also used in the unimodular version of GR), the traceless Brans-Dicke equations are set forth, and it is shown that after imposing the conservation of the energy-momentum tensor, the usual Brans-Dicke equations are recovered with an additional cosmological constant type term. However, if the conservation of the energy-momentum tensor is not imposed, since it is not dictated by the TDiff, the equivalence disappears, as it happens in the GR case. Hence, the overall relations between
a unimodular Brans-Dicke (UBD) theory repeats the same theoretical pattern as that observed in the unimodular GR (UGR) case.

Again, we prove that in the UBD  background case the vacuum solutions are the same as the the BD solutions in presence of a
cosmological constant. However, at perturbative level, this equivalence is lost, as it is shown explicitly. This is due to the
unimodular condition which must be taken into account. In particular, an unstable solution appears, while in the BD case
only stable solutions are present in the vacuum case, with or without the cosmological constant.

We organize this article as follows: In the next section, we revise how to implement the unimodular version of the GR
theory using a constrained system. This procedure is repeated in the Brans-Dicke context in section 3. The cosmological
scenarios in GR, BD and UBD cases are analyzed in section 4, while in section 5 a perturbative study is carried out. Our
conclusions are presented in section 6.

\section{Unimodular structure in GR}

Prior to describing the implementation of the unimodular constraint in the Brans-Dicke scalar-tensor theory, we revise
briefly one of its possible implementations in the context of General Relativity theory through the addition of the constraint
in the Einstein-Hilbert action using the Lagrange multiplier technique. We follow closely the approach of Ref. \cite{brand}.

Let us consider the action,
\begin{eqnarray}
{\cal S} = \int d^4x\biggr\{\sqrt{-g}R - \chi(\sqrt{-g} - \xi)\biggl\} + \int d^4 x \sqrt{-g}{\cal L}_m.
\end{eqnarray}
In this expression, $\chi$ is a Lagrange multiplier, and $\xi$ is in principle an external field introduced in order to give flexibility in the choice of the coordinate system. As a special case, $\xi$ may be consider a number, which implies to choose a specific coordinate system. The variation with respect to the metric leads to,
\begin{eqnarray}
\label{erg1}
R_{\mu\nu} - \frac{1}{2}g_{\mu\nu}R + \frac{\chi}{2}g_{\mu\nu} = 8\pi GT_{\mu\nu}.
\end{eqnarray}
The variation with respect to $\chi$ leads to the unimodular constraint:
\begin{eqnarray}
\label{vin-rg-1}
\xi = \sqrt{-g}.
\end{eqnarray}

The trace of (\ref{erg1}) implies:
\begin{eqnarray}
\chi = \frac{R}{2} + 8\pi G\frac{T}{2}.
\end{eqnarray}
 Inserting this result in (\ref{erg1}), we obtain the unimodular field equations:
 \begin{eqnarray}
 \label{erg2}
 R_{\mu\nu} - \frac{1}{4}g_{\mu\nu}R = 8\pi G\biggr(T_{\mu\nu} - \frac{1}{4}g_{\mu\nu}T\biggl).
 \end{eqnarray}
 The theory is now invariant by a restricted class of diffeomorphisms, called transverse diffeomorphism \cite{trans}.
 
 The Bianchi identities imply that, in general, the usual energy-momentum tensor conservation must be generalized as follows
 \begin{eqnarray}
 \label{brg1}
 \frac{R^{;\nu}}{4} = 8\pi G\biggr({T^{\mu\nu}}_{;\mu} - \frac{1}{4}T^{;\nu}\biggl).
 \end{eqnarray}

 If the usual energy-momentum tensor conservation is imposed (which is, in the unimodular context, a choice in opposition to GR theory),
 \begin{eqnarray}
 \label{cons-rg-1}
 {T^{\mu\nu}}_{;\mu} = 0,
 \end{eqnarray}
 equation (\ref{brg1}) becomes,
 \begin{eqnarray}
 \label{brg2}
 \frac{R^{;\nu}}{4} = -2 \pi GT^{;\nu}.
 \end{eqnarray}
Equation (\ref{brg2}) may be integrated leading to,
\begin{eqnarray}
\label{ic1}
R = - 2\pi GT + 4\Lambda,
\end{eqnarray}
where $\Lambda$ is an integration constant which may be associated to the cosmological constant. 
Inserting the relation (\ref{ic1}) in (\ref{erg1}), the resulting equations are,
\begin{eqnarray}
 \label{erg3}
 R_{\mu\nu} - \frac{1}{2}g_{\mu\nu}R = 8\pi GT_{\mu\nu} + g_{\mu\nu}\Lambda.
 \end{eqnarray}
 The GR equations are recovered, but with a cosmological constant which was absent in the original structure. Note that
it emerges naturally as a constant of integration.  
 
\section{Unimodular Brans-Dicke}

For the Brans-Dicke theory we will consider a structure similar to that present previously. The Lagrangian leading to the original BD theory is given by \cite{bd},
\begin{eqnarray}
\label{bd}
{\cal S} = \int d^4x\bigg\{\sqrt{-g}\biggr[\phi R - \omega\frac{\phi_{;\rho}\phi^{;\rho}}{\phi} \biggl] + {\cal L}_m\biggl\}.
\end{eqnarray}
The unimodular constrained Lagrangian is then given by,
\begin{eqnarray}
\label{cbd}
{\cal S} = \int d^4x\bigg\{\sqrt{-g}\biggr[\phi R - \omega\frac{\phi_{;\rho}\phi^{;\rho}}{\phi}  + 2V(\phi)\biggl] - \phi\chi(\sqrt{-g} - \xi) + {\cal L}_m\biggl\}.
\end{eqnarray}
The potential $V(\phi)$ has been introduced in a convenient form in order to keep contact with $f(R)$ class of theories \cite{f(R)}. In the usual Brans-Dicke theory, $V(\phi) = 0$.

The variational principle leads to the equations:
\begin{eqnarray}
\label{bde1-a}
R_{\mu\nu} - \frac{1}{2}g_{\mu\nu}R &=& \frac{8\pi}{\phi}T_{\mu\nu} + \frac{\omega}{\phi^2}(\phi_{;\mu}\phi_{;\nu} - \frac{1}{2}g_{\mu\nu}\phi_{;\rho}\phi^{;\rho}) + g_{\mu\nu}\frac{V(\phi)}{\phi}\nonumber\\
&+& \frac{1}{\phi}(\phi_{;\mu\nu} - g_{\mu\nu}\Box\phi) - g_{\mu\nu}\frac{\chi}{2},\\
\label{bde2-a}
2\omega\Box\phi &=& \omega\frac{\phi_{;\rho}\phi^{;\rho}}{\phi} - \phi R + \chi(\sqrt{-g} - \xi) - 2\phi V_\phi(\phi),\\
\label{bde3-a}
\sqrt{-g} &=& \xi.
\end{eqnarray}
The subscript $\phi$ in the potential term indicates derivative with respect to the scalar field.

Inserting the constraint (\ref{bde3-a}) in (\ref{bde2-a}), it comes out,
\begin{eqnarray}
\label{bde2-b}
2\omega\Box\phi &=& \omega\frac{\phi_{;\rho}\phi^{;\rho}}{\phi} - \phi R - 2\phi V_\phi(\phi).
\end{eqnarray}
The trace of (\ref{bde1-a}) leads to
\begin{eqnarray}
R = - \frac{8\pi}{\phi}T + \omega\frac{\phi_{;\rho}\phi^{;\rho}}{\phi^2} - 4\frac{V(\phi)}{\phi}+ 3\frac{\Box\phi}{\phi}+ 2\chi.
\end{eqnarray}
Using the above expression to eliminate $\chi$ from the field equations, we obtain
\begin{eqnarray}
\label{bde1-f1}
R_{\mu\nu} - \frac{1}{4}g_{\mu\nu}R &=& \frac{8\pi}{\phi}\biggr(T_{\mu\nu} - \frac{1}{4}g_{\mu\nu}T\biggl) + \frac{\omega}{\phi^2}(\phi_{;\mu}\phi_{;\nu} - \frac{1}{4}g_{\mu\nu}\phi_{;\rho}\phi^{;\rho})\nonumber\\
&+& \frac{1}{\phi}\biggr(\phi_{;\mu\nu} - \frac{1}{4}g_{\mu\nu}\Box\phi\biggl),\\
\label{bde18-f}
\Box\phi &=& \frac{1}{2}\frac{\phi_{;\rho}\phi^{;\rho}}{\phi} - \frac{\phi}{2\omega}R - \frac{\phi}{\omega}V_\phi(\phi).
\end{eqnarray}
Remark that the potential has disappeared from the field equations (\ref{bde1-f1}). The Bianchi identities i.e., the vanishing divergence of the Einstein tensor hidden in the left hand side on (\ref{bde18-f}), will imply in
\begin{eqnarray}
\label{bbd1}
(\phi R)^{;\nu} = \omega\biggr(\frac{\phi_{;\rho}\phi^{;\rho}}{\phi}\biggl)^{;\nu} + 32\pi \biggr({T^{\mu\nu}}_{;\mu} - \frac{1}{4}T^{;\nu}\biggl) + 3(\Box\phi)^{;\nu} - 4V^{;\nu}.
\end{eqnarray}

In the case where energy-momentum tensor conservation is imposed, 
\begin{eqnarray}
{T^{\mu\nu}}_{;\mu} = 0,
\end{eqnarray}
we obtain the relation,
\begin{eqnarray}
R = \omega\frac{\phi_{;\rho}\phi^{;\rho}}{\phi^2} - \frac{8\pi T}{\phi} + 3\frac{\Box\phi}{\phi}  - 4 \frac{V(\phi)}{\phi} - 4\frac{\Lambda}{\phi},
\end{eqnarray}
and inserting this relation in the field equations, we obtain,
\begin{eqnarray}
\label{bde1-f}
R_{\mu\nu} - \frac{1}{2}g_{\mu\nu}R &=& \frac{8\pi}{\phi}T_{\mu\nu}+ \frac{\omega}{\phi^2}(\phi_{;\mu}\phi_{;\nu} - \frac{1}{2}g_{\mu\nu}\phi_{;\rho}\phi^{;\rho})\nonumber\\
&+& \frac{1}{\phi}(\phi_{;\mu\nu} - g_{\mu\nu}\Box\phi) + g_{\mu\nu}\frac{V}{\phi} +  g_{\mu\nu}\frac{\Lambda}{\phi},\\
\label{bde2-f}
\Box\phi &=& \frac{8\pi}{3 + 2\omega}T + \frac{2}{3 + 2\omega} \biggr(2V(\phi) - \phi V_\phi(\phi)\biggl) + \frac{4}{3 + 2\omega}\Lambda.
\end{eqnarray}

On the other hand, without the conservation of the energy-momentum tensor, the equations are:
\begin{eqnarray}
\label{bde1-ff}
R_{\mu\nu} - \frac{1}{4}g_{\mu\nu}R &=& \frac{8\pi}{\phi}\biggr(T_{\mu\nu} - \frac{1}{4}g_{\mu\nu}T\biggl) + \frac{\omega}{\phi^2}(\phi_{;\mu}\phi_{;\nu} - \frac{1}{4}g_{\mu\nu}\phi_{;\rho}\phi^{;\rho})\nonumber\\
&+& \frac{1}{\phi}(\phi_{;\mu\nu} - \frac{1}{4}g_{\mu\nu}\Box\phi),\\
\label{bde2-ff}
\Box\phi &=& \frac{1}{2}\frac{\phi_{;\rho}\phi^{;\rho}}{\phi} - \frac{\phi}{2\omega}R - \frac{\phi}{\omega}V_\phi(\phi),\\
\label{bde3-ff}
(\phi R)^{;\nu} &=& \omega\biggr(\frac{\phi_{;\rho}\phi^{;\rho}}{\phi}\biggl)^{;\nu} + 32\pi \biggr({T^{\mu\nu}}_{;\mu} - \frac{1}{4}T^{;\nu}\biggl) + 3(\Box\phi)^{;\nu} - 4V^{;\nu}.
\end{eqnarray}

Equations (\ref{bde1-f}-\ref{bde3-ff}) represent the unimodular version of the BD theory.

\section{Cosmology}

We will explore now some consequences of the BD unimodular theory given by (\ref{bde1-ff}-\ref{bde3-ff}). We will concentrate on the cosmological flat model defined by the metric, 
\begin{eqnarray}
\label{metrica}
ds^2 = dt^2 - a^2(dx^2 + dy^2 + dz^2),
\end{eqnarray}
a$\equiv a(t)$ is the scale factor. But, before analyzing the BD unimodular theory, we revise briefly some properties of the RG unimodular theory for cosmology.

\subsection{Unimodular RG}

Using the metric (\ref{metrica}), the equations for the unimodular theory in the RG context become:
\begin{eqnarray}
\label{urg1}
\dot H &=& - 4\pi G\bar\rho,\\
\label{urg2}
\ddot H + 4 H\dot H &=& - 4\pi G(\dot{\bar\rho} + 4H\bar\rho),
\end{eqnarray}
with the definitions,
\begin{eqnarray}
H &=& \frac{\dot a}{a},\\
\bar\rho &=& \rho + p,
\end{eqnarray}
where dot represents time derivative. Remark that the equations are sensitive only to the combination $\bar\rho = \rho + p$. Hence, the "vacuum" condition $\bar\rho = 0$ can lead to the Minkowski or de Sitter space-times.
Moreover, if we substitute (\ref{urg1}) into (\ref{urg2}) we obtain an identity. Hence, the two equations have the same content essentially and  we have only one equation for two variables, $H$ and $\bar\rho$: the system is underdetermined for $\bar\rho \neq 0$.
This implies that in order to obtain non-vacuum solutions, we need to impose an ansatz to close the system of equations.

The vacuum case in UGR is essentially trivial. For this reason, we just sketch the behaviour in the non-vacuum case with a specific ansatz to close the system of equations, which corresponds to the radiative universe.

It seems natural, due to the traceless character of the field equations, to assume that the equations satisfy the usual radiative solution:
\begin{eqnarray}
H &=& \frac{1}{2t},\\
\bar\rho &=& \frac{\bar\rho_0}{a^{4}}.
\end{eqnarray}
This happens independently of the equation of state connecting $p$ and $\rho$. In fact, since the system is underdetermined, we can impose any behavior for the Hubble function or, equivalently, for the matter component. However, the radiative behavior, with $\bar\rho \propto a^{-4}$ is well justified for a system composed only of radiation: for a gas of photons, the density of photons decreases, as the universe expands, as the cube of the scale factor and the wavelength of each photon is stretched by the expansion, leading to the
law for $\bar\rho$ given above. 

The previous ansatz has been studied in detail in Ref. \cite{velten}. 
There is a first integral given by,
\begin{eqnarray}
\label{cec}
\dot H + 2H^2 =  \frac{2}{3}\Lambda_{\rm U}, \quad \Lambda_{\rm U}  = \mbox{constant}.
\end{eqnarray}
The integration constant, which we have called $\Lambda_U$, makes the unimodular cosmological scenario essentially identical to the GR radiative model in presence of a cosmological constant. As shall show bellow, it is convenient to introduce the factor $2/3$.

From (\ref{cec}) we have three possibilities:
\begin{eqnarray}
\Lambda_{\rm U} < 0 \quad &\rightarrow& \quad a = a_0\sin^{1/2}\sqrt{-\frac{4 \Lambda_{\rm U}}{3}} t,\\
\Lambda_{\rm U} = 0 \quad &\rightarrow& \quad a = a_0 t^{1/2},\\
\Lambda_{\rm U} > 0 \quad &\rightarrow& \quad a = a_0\sinh^{1/2}\sqrt{ \frac{4 \Lambda_{\rm U}}{3}}t. \label{apositiveLambda}
\end{eqnarray}

In Ref. \cite{velten} it was shown that a viable cosmological model is achieved for values of $\Lambda_U \lesssim 1$ with an age of the universe compatible with observations i.e., the universe can be older than the age of 
globular clusters. Also, the late time dynamics of this model in quite different from the standard cosmology. The model transits from a radiative evolution directly to a de Sitter phase. There is no matter dominated epoch. However, the analysis of the evolution for scalar density perturbations reveals a stronger agglomeration of matter in comparison with the standard Einstein-de Sitter. Then, even in such non-trivial background, structures can reach the nonlinear regime

\subsection{Brans-Dicke solutions}

At this stage, it is important to recall the vacuum Brans-Dicke solutions in presence or not of a cosmological constant.
The equations are the following.
\begin{eqnarray}
3H^2 &=& \frac{8\pi}{\phi}\rho + \frac{\omega}{2}\frac{\dot\phi^2}{\phi^2} - 3H\frac{\dot\phi}{\phi} + \frac{\Lambda}{\phi}, \\
\frac{\ddot\phi}{\phi} + 3H\frac{\dot\phi}{\phi} &=& \frac{4\Lambda + 8\pi(\rho - 3 p)}{(3 + 2\omega)\phi}.
\end{eqnarray}
For the vacuum case, we set $\rho = p = 0$, leading to,
\begin{eqnarray}
\label{ce1}
3H^2 &=& \frac{\omega}{2}\frac{\dot\phi^2}{\phi^2} - 3H\frac{\dot\phi}{\phi} + \frac{\Lambda}{\phi}, \\
\label{ce2}
\frac{\ddot\phi}{\phi} + 3H\frac{\dot\phi}{\phi} &=& \frac{4\Lambda}{(3 + 2\omega)\phi}.
\end{eqnarray}
These equations admit simple solutions which may contain or not the cosmological term. They are the following.
\begin{enumerate}
\item Vacuum solutions with $\Lambda = 0$.
\begin{enumerate}
\item Power law solutions appears when $\omega \neq - 4/3$.
\begin{eqnarray}
a &=& a_0t^\frac{-(1 + \omega) \pm \sqrt{1 + 2\omega/3}}{4 + 3\omega},\\
\phi &=& \phi_0 t^\frac{1 \mp \sqrt{1 + 2\omega/3}}{3(4 + 3\omega)}.
\end{eqnarray}
\item Exponential solutions for $\omega = - 4/3$.
\begin{eqnarray}
a &=& a_0 e^{Ht},\\
\phi &=& \phi_0 e^{-3H},
\end{eqnarray}
with $H = $ constant.
\end{enumerate}
\item Vacuum solutions with $\Lambda \neq 0$.
\begin{enumerate}
\item Power law solutions for any value of $\omega$ such that,
\begin{eqnarray}
a &=& a_0t^{\omega + 1/2},\\
\phi &=& \phi_0 t^2.
\end{eqnarray}
Remark that, for $\omega = - 1/2$, there is a solution describing a static universe but with an evolving gravitational coupling.
\end{enumerate}
\end{enumerate}
In this analysis, we have excluded the case $\omega = - 3/2$ for which the theory is conformally equivalent to GR.

\subsection{Unimodular Brans-Dicke cosmology}

Let us turn now to the BD unimodular context given by (\ref{bde1-ff}-\ref{bde3-ff}).
The equations of motion describing the corresponding cosmological model are the following.
\begin{eqnarray}
\label{uc1}
\dot H &=& - \frac{4\pi}{\phi}(\rho + p) - \frac{\omega}{2}\frac{\dot\phi^2}{\phi^2} - \frac{1}{2}\biggr(\frac{\ddot\phi}{\phi} - H\frac{\dot\phi}{\phi}\biggl),\\
\label{uc2}
\frac{\ddot \phi}{\phi} + 3H\frac{\dot\phi}{\phi} &=& \frac{1}{2}\frac{\dot \phi^2}{\phi^2} + \frac{3}{\omega}(\dot H + 2H^2).
\end{eqnarray}

We consider the possibility of exponential solutions, as it happens in the Brans-Dicke case. Let us suppose the solutions under the form,
\begin{eqnarray}
a &=& a_0e^{Ht}, \quad H = cte,\\
\phi &=& \phi_0e^{st}, \quad s = cte.
\end{eqnarray}
Using (\ref{uc1},\ref{uc2}), the resulting relations for  $s$ are,
\begin{eqnarray}
H &=& (1 + \omega)s,\\
s^2 + 6Hs - \frac{12}{\omega}H^2 &=& 0.
\end{eqnarray}
If $\omega = -1$, it comes out that $H = s = 0$, that is, the Minkowski solution. Remember that $\omega = -1 $, in the original Brans-Dicke context corresponds to the low energy string configuration.
If $\omega \neq -1$, those relations imply that $s$ is given by,
\begin{eqnarray}
1 = \biggr(- 3 \pm \sqrt{9 + \frac{12}{\omega}}\biggl)(1 + \omega).
\end{eqnarray}
There are two possible solutions: $\omega = - 4/3$, in agreement with the usual Brans-Dicke result; $\omega = - 3/2$, a case where the BR theory becomes conformal equivalent to GR. 

Now we can look for power law solutions via the power-law ansatz
\begin{eqnarray}
a &=& a_0t^r,\\
\phi &=& \phi_0t^s,
\end{eqnarray}
we find the relations,
\begin{eqnarray}
(s + 2)r-s+(1 + \omega)s^2 &=& 0,\\
(s^2 - 2s)\omega + 6(\omega s + 1)r - 12r^2 &=& 0.
\end{eqnarray}
It is direct to verify that the BD corresponding solutions ($r = \omega + 1/2$ and $s = 2$) are also solutions here. But is there any other non trivial solution?

We can combine the above relations to obtain a single algebraic equation for $s$.
\begin{eqnarray}
(12 + 17\omega + 6\omega^2)s^3 - (30 + 38\omega + 12\omega^2)s^2 + (6 + 4\omega)s + 2(6 + 4\omega) = 0.
\end{eqnarray}
Since $s = 2$ is solution, we can reduce the order of the equation for $s$:
\begin{eqnarray}
(12 + 17\omega + 6\omega^2)s^2 - (6 + 4\omega)s - (6 + 4\omega) = 0.
\end{eqnarray}
This equation can be rewritten as,
\begin{eqnarray}
\biggr(2 + \frac{3}{2}\omega\biggl)s^2 - s - 1 = 0,
\end{eqnarray}
with the solutions,
\begin{eqnarray}
s = \frac{1 \pm \sqrt{1 + \frac{2}{3}\omega}}{3(4 + 3\omega)}.
\end{eqnarray}
corresponding to the vacuum Brans-Dicke solutions without the cosmological constant. 

In brief, the vacuum background solutions are identical to the usual vacuum solutions of the BD theory with and without a cosmological constant, similarly to what happens in the GR unimodular case.

\section{Perturbations}

We turn now to the perturbative analysis. One main point is that the unimodular condition $\sqrt{-g} = \xi$ combined with the linear perturbations such that $g_{\mu\nu} \rightarrow g_{\mu\nu} + h_{\mu\nu}$, impliy that,
\begin{eqnarray}
h_{kk} = 0.
\end{eqnarray}
As discussed in \cite{velten,brand}, this condition, which is also a consequence of the transverse diffeomorphism on which the unimodular theory is based, leads to the impossibility to use the newtonian gauge. The synchronous gauge can not be used either if the energy-momentum tensor conserves, but if the generalized conservation laws (\ref{brg1}) or (\ref{bbd1}) are used, it is consistent to use the synchronous gauge condition. The gauge invariant formalism can be used in any case. We remark that, as it will be seen below, the use of the synchronous gauge leads to a second order differential equation for the perturbations. This implies that, in opposition to what happens in the GR or BD usual theories, there is no residual gauge mode, a consequence of the the TDiff obeyed by the unimodular theories as it can be verified explicitly.

From now on, we will restricted to the synchronous gauge in the unimodular context, implying,
\begin{eqnarray}
h_{\mu0} = 0, \quad h_{kk} = 0.
\end{eqnarray}
Under these conditions the perturbations of the Ricci tensor are (which can be obtained directly by imposing the unimodular constraints in the expressions found, for example, in \cite{weinberg}),
\begin{eqnarray}
\delta R_{00} &=& 0, \\
\delta R_{0i} &=& - \frac{\dot f_i}{2},\\
\delta R_{ij} &=& \frac{1}{2}\biggr(\nabla^2 f_{ij} - f_{i,j} - f_{j,i}\biggl) - \frac{\ddot h_{ij}}{2} + \frac{H}{2}\dot h_{ij} - 2H^2h_{ij},\\
\delta R &=& \frac{f}{a^2}.
\end{eqnarray}
with the definitions,
\begin{eqnarray}
f_{ij} = \frac{h_{ij}}{a^2}, \quad f_i = \frac{h_{ki,k}}{a^2}, \quad f = \frac{h_{kl,k,l}}{a^2}.
\end{eqnarray}

Perturbing the unimodular gravitational tensor, $E_{\mu\nu}= R_{\mu\nu} - \frac{1}{4}g_{\mu\nu}R$ and using the previous relations, we find,
\begin{eqnarray}
\delta E_{00} &=& - \frac{1}{4}\frac{f}{a^2},\\
\delta E_{0i} &=& - \frac{\dot f_i}{2},\\
\delta E_{ij} &=& \frac{1}{2}\biggr[\nabla^2 f_{ij} - f_{i,j} - f_{j,i} + \delta_{ij}\frac{f}{2}\biggl] 
- \frac{\ddot h_{ij}}{2} + \frac{H}{2}\dot h_{ij} + 
\biggr(\frac{3}{2}\dot H + H^2\biggl)h_{ij}.
\end{eqnarray}

Perturbing both sides of the field equations, we obtain the following expressions driving the perturbations:

\begin{eqnarray}
\delta E_{\mu\nu} &=& \frac{8\pi}{\phi}\biggr\{\delta T_{\mu\nu} - \frac{1}{4}\biggr(h_{\mu\nu}T + g_{\mu\nu}\delta T\biggl)\biggl\} \nonumber\\
&+& \frac{\omega}{\phi^2}\biggr\{\delta\phi_{;\mu}\phi_{;\nu} + \phi_{;\mu}\delta\phi_{;\nu} - \frac{1}{4}h_{\mu\nu}\phi_{;\rho}\phi^{;\rho} - \frac{1}{4}g_{\mu\nu}\biggr(- h^{\rho\sigma}\phi_{;\rho}\phi_{;\sigma} + 2\delta\phi_{\rho}\phi^{\rho}\biggl)\biggl\}\nonumber\\
&+& \frac{1}{\phi}\biggr\{\delta(\phi_{;\mu;\nu}) - \frac{1}{4}\biggr(h_{\mu\nu}\Box\phi + g_{\mu\nu}\delta(\Box\phi)\biggl)\biggl\}\nonumber\\
&-& \frac{\delta\phi}{\phi}E_{\mu\nu} - \omega\frac{\delta\phi}{\phi}\biggr\{\phi_{;\mu;\nu} - \frac{1}{4}g_{\mu\nu}\phi_{;\rho}\phi^{;\rho}\biggl\} \,,
\end{eqnarray}
where $\delta \phi$ is the scalar field $\phi$ perturbation. From this expression, we obtain the following equations for its components.

\begin{itemize}
\item $\mu = \nu = 0$:

\begin{eqnarray} \label{oo}
\frac{f}{a^2} &=& - \frac{24\pi}{\phi}(\delta\rho + \delta p) - 3\frac{\delta\ddot\phi}{\phi} + 3\biggr(H - 2\omega\frac{\dot\phi}{\phi}\biggl)\frac{\delta\dot\phi}{\phi} \nonumber\\
&-& 3\biggr(2\dot H - \omega\frac{\dot\phi^2}{\phi^2}\bigg)\frac{\delta\phi}{\phi} - \frac{\nabla^2\delta\phi}{a^2\phi}.
\end{eqnarray}

\item $\mu = 0, \nu = i$:
\begin{eqnarray}
\label{pe1}
\dot f_i = - \frac{16 \pi}{\phi}(\rho + p)\delta u_i - 2\omega\frac{\dot\phi}{\phi^2}\delta\phi_{,i} - \frac{2}{\phi}\biggr(\delta\dot\phi_{,i} - H\delta\phi_{,i}\biggl)
\end{eqnarray}

\item $\mu = i, \nu = j$:

\begin{eqnarray}
\label{ij}
&& \frac{1}{a^2}\biggr(\nabla^2 h_{ij} - h_{ik,j,k} - h_{kj,i,k}\biggl) + \frac{f}{2}\delta_{ij} - \ddot h_{ij} +\biggr(H - \frac{\dot\phi}{\phi}\biggl)\dot h_{ij}  + \biggr(2\dot H + 2 H^2 +
2H\frac{\dot\phi}{\phi} \biggl)h_{ij} \nonumber\\
 &=& \frac{4\pi}{\phi}(\delta\rho + \delta p)a^2\delta_{ij}\nonumber\\
 &+& \frac{a^2}{2}\biggr[\frac{\delta\ddot\phi}{\phi} + \biggr(2\omega\frac{\dot\phi}{\phi} - H\biggl)\frac{\delta\dot\phi}{\phi}
 + \biggr(2\dot H - \omega\frac{\dot\phi^2}{\phi^2}\biggl)\frac{\delta\phi}{\phi}\biggl]\delta_{ij} + 2\frac{\delta\phi_{,i,j}}{\phi} - \frac{1}{2}\frac{\nabla \delta\phi}{\phi}\delta_{ij}.
 \end{eqnarray}
  The perturbation of the Klein-Gordon equation is,
 \begin{eqnarray}
 \label{pkg}
 \delta\ddot\phi + \biggr(3H - \frac{\dot\phi}{\phi}\biggl)\delta\dot\phi + \biggr[\frac{1}{2}\frac{\dot\phi^2}{\phi^2} - 
 \frac{3}{\omega}(\dot H + 2H^2)\biggl]\delta\phi - \frac{\nabla^2}{a^2}\delta\phi = - \frac{\phi}{2\omega}\frac{f}{a^2}.
 \end{eqnarray} 
 \end{itemize}
Eqs. (\ref{oo})-(\ref{pkg}) govern scalar, vectorial and tensor perturbations dynamics in unimodular cosmology. We will concentrate on the tensorial and scalar perturbations

\subsection{Gravitational waves}

From (\ref{ij}) we can obtain the equation for the tensorial mode, which reads, after performing a Fourier decomposition,
\begin{eqnarray}
\label{gw}
\ddot h_{ij} - \biggr(H - \frac{\dot\phi}{\phi}\biggl)\dot h_{ij} + \bigg[\frac{k^2}{a^2} - \bigg(2\dot H + 2 H^2 + 2H\frac{\dot\phi}{\phi}\biggl)\biggl]h_{ij} = 0.
\end{eqnarray} 
This is the same equation for gravitational waves in the original Brans-Dicke theory, repeting what happens in the UGR case \cite{gw}.
After passing to the conformal time, which is related to the cosmic time by $\eta \propto t^{-\omega + 1/2}$, the equation can be rewritten as a Bessel equation with the solution,
\begin{eqnarray}
h_{ij} = \epsilon_{ij}C_\pm \eta^{-1/2}J_{\pm\nu}(k\eta)
\end{eqnarray}
where $\epsilon_{ij}$ is the constant polarization tensor, $C_\pm$ are constants, and $\nu$ is the order of the Bessel function,

\begin{eqnarray}
\nu = \frac{(6\omega+5)}{2|1 - 2\omega|}.
\end{eqnarray}
In the limit $\omega \rightarrow \infty$ the GR result is re-obtained.

For $\omega = 1/2$ the gravitational wave equation becomes a Euler-type equation with the solution,
\begin{eqnarray}
h_{ij} = \epsilon_{ij}c_\pm t^\frac{ -1 \pm \sqrt{1 - 4\bar k}}{2},
\end{eqnarray}
with $\bar k = k/a_0$.

\subsection{Vacuum scalar perturbations}

We call vacuum perturbations the case for $\delta\rho =\delta p = 0$, which may include or not a cosmological constant.
Imposing this condition,  and after combining equations (\ref{oo}) and (\ref{pkg}), we obtain a single equation for $\delta\phi$: 
\begin{eqnarray}
\label{master2}
(3 - 2\omega)\delta\ddot\phi - \biggr[3(1 + 2\omega)H - 8\omega\frac{\dot\phi}{\phi}\biggl]\delta\dot\phi \nonumber\\
+ \biggr[12(\dot H + H^2) - 4\omega\frac{\dot\phi^2}{\phi^2}\biggl]\delta\phi + \frac{1 + 2\omega}{a^2}\nabla^2\delta\phi = 0.
\end{eqnarray}

Even without integrating this equation we can have an information about the relative sign of the two time derivative leading term and the laplacian, which must be negative in order to be stable. This is achieved for $\omega < - 1/2$ or $\omega > 3/2$. In the interval $- 1/2 < \omega <  3/2$ the solution is unstable. Remember that $\omega = - 1/2$ is the case of a static universe.

Inserting the background solution in (\ref{master2}), and after performing a Fourier decomposition, we obtain,
\begin{eqnarray}
\delta\ddot\phi + r\frac{\delta\dot\phi}{t} + \biggr(\bar c_s^2\frac{k^2}{t^{2\omega + 1}} - \frac{s}{t^2}\biggl)\delta\phi = 0,
\end{eqnarray}
with,
\begin{eqnarray}
r &=& - \frac{(1 - 6\omega)}{2},\\
s &=& 1 + 6\omega,\\
\bar c_s^2 &=& - \frac{1 + 2\omega}{3 - 2\omega}\frac{1}{a_0^2}.
\end{eqnarray}

Performing the variable transformation,
\begin{eqnarray}
x = t^p, \quad p = \frac{1 - 2\omega}{2},
\end{eqnarray}
the equation takes the form,
\begin{eqnarray}
\delta\phi'' - 2\frac{\delta\phi'}{x} + \biggr(\frac{\bar c_s^2k^2}{p^2} - \frac{\bar s}{x^2}\biggl)\delta\phi = 0,
\end{eqnarray}
with $\bar s = s/p^2$.

With the transformation $\delta\phi = x^{3/2}\lambda$, we obtain the equation
\begin{eqnarray}
\label{bessel}
\lambda'' + \frac{\lambda'}{x} + \biggr(\frac{\bar c_s^2k^2}{p^2} - \frac{\nu^2}{x^2}\biggl)\lambda = 0,
\end{eqnarray}
with $\nu^2 = \bar s + 9/4$, which can be simplified to give,
\begin{eqnarray}
\nu = \pm \frac{(5 + 6\omega)}{2(1 - 2\omega)}.
\end{eqnarray}
 Equation (\ref{bessel}) is a Bessel equation of order $\nu$.
The final solution is,
\begin{eqnarray}
\delta\phi = x^{3/2}\biggr\{c_1 J_\nu(\bar k x) + c_2 J_{-\nu}(\bar k x)\biggl\},
\end{eqnarray}
where $c_{1,2}$ are integration constants, and $\bar k = \sqrt{\bar c_s^2} k/p$. When $\bar c_s^2 < 0$, the solutions are expressed in terms of modified Bessel functions:
\begin{eqnarray}
\delta\phi = x^{3/2}\biggr\{c_1 K_\nu(|\bar k| x) + c_2 I_\nu(|\bar k |x)\biggl\}.
\end{eqnarray}

We inspect now the two particular cases, $\omega = -1/2$ and $\omega= 3/2$. For $\omega = - 1/2$ the universe is static, but the scalar field evolves as $\phi \propto t^2$. The perturbative equation reads,
\begin{eqnarray}
\delta\ddot\phi - 2\frac{\delta\dot\phi}{t} + 2\frac{\delta\phi}{t^2} = 0.
\end{eqnarray}
This is an Euler type equation with solutions,
\begin{eqnarray}
\delta\phi = c_1 t^2 + c_2 t, \quad \quad c_{1,2} = \mbox{constant}.
\end{eqnarray}
This static universe is stable since at most the perturbations evolve as the background. This is an example of a static and stable universe.

For $\omega = 3/2$, the scale factor and the scalar field both evolves as $t^2$. The perturbed equation reduces to,
\begin{eqnarray}
\label{master}
\nabla^2\delta\phi = 0.
\end{eqnarray}
This implies that the perturbations are homogeneous. Since the function $f$ contains spatial derivatives, it must be zero. This leads to a unique equation for $\delta\phi$, which reads,
\begin{eqnarray}
\delta\ddot\phi + 4\frac{\delta\dot\phi}{t}  - 10 \frac{\delta\phi}{t^2} = 0.
\end{eqnarray}
This is an Euler type equation with solutions,
\begin{eqnarray}
\delta\phi = d_1 t^2 + d_2 t^{-5}.
\end{eqnarray}
The perturbations (which has no space dependence) evolve at most at the same rate as the background implying stability.

The comparison with the results obtained in the perturbative analysis of the Brans-Dicke theory, obtained in Ref. \cite{plinio}, is not so direct since the synchronous coordinate condition (which also has been used in that reference) displays certain special features here due to the unimodular condition. But, at small scale the usual Brans-Dicke case displays no instabilities since the speed of sound is always positive, while in the present version there is a clear instability in the interval $- 1/2 < \omega < 3/2$. Moreover, the particularities at $\omega = -1/2, 3/2$ do not appear in the usual Brans-Dicke case.

\section{Conclusions}

We have constructed the unimodular version of the Brans-Dicke theory using an action with its constraint expressed by a Lagrangian multiplier. In this way, traceless field equations have been obtained, complemented by the unimodular condition on the determinant of the metric. In such formulation, this constraint has been presented with the help of an external field in such a way
that the coordinate system is not fixed as in the case when the determinant of the metric is fixed equal to one. As in the unimodular version of General Relativity, the usual conservation laws are replaced by a more complex relation obtained by the employment of the Bianchi identities. If the usual conservation laws are imposed, the Brans-Dicke equations in presence of a cosmological constant are recovered. But, this is not mandatory, and we have explored the possibility of having generalized conservation laws, what may in principle break the correspondence with the traditional Brans-Dicke theory.

We have paid attention to the cosmological set up. When we do not consider the usual conservation law, all equations are sensitive only to the combination $\rho + p$, that is, the enthalpy of the total matter sector. The same feature appears in the UGR case as considered in \cite{velten}. This has an interesting consequence: the cases $\rho = p = 0$ and $\rho = - p$ are equivalent since they lead to the vacuum solutions in presence of an integration constant which can be identified with the cosmological constant (which may be fixed equal zero, of course). The resulting solutions are the same as the usual BD vacuum solutions either with or without a cosmological constant.

However, this equivalence is broken at perturbative level mainly due to the unimodular condition. The latter condition forbids the use of the synchronous and newtonian gauges if the
divergence of the energy-momentum tensor is fixed equal to zero. If not, in the nonconservative case (see Ref. \cite{Velten:2021xxw}
 for a review of nonconservative theories), the synchronous gauge condition is allowed (the newtonian one remains forbidden). The vacuum perturbed equations can be solved analytically and they are not the same as in the usual Brans-Dicke theory. In particular, the perturbations reveal instabilities in the small scale limit if the BD parameter is in the interval $- 1/2 < \omega < 3/2$. For all other values of $\omega$, the perturbations are stable.

The introduction of more general forms of matter shares the same problem that in UGR: the resulting system of equation is undetermined and an extra assumption must be introduced in order to obtain a self-consistent scenario. We intend to investigate this problem in the future.
\noindent
\section*{Acknowledgments}
We thank FAPES, CAPES and CNPq for partial financial support. JCF thanks ITP of the Universität
Heidelberg and the LPTMC of the Sorbonne-Université and CNRS  for the kind hospitality during the elaboration of this work. WSHR thanks FAPES (Brazil) for financial support (PRONEM No 503/2020). 


\end{document}